# Interlayer interactions in one-dimensional van der Waals moiré superlattices

Sihan Zhao[1*], Ryo Kitaura[2*], Pilkyung Moon[3,4], Mikito Koshino[5], Feng Wang[6,7,8*]

1. Interdisciplinary Center for Quantum Information, State Key Laboratory of Silicon Materials, and Zhejiang Province Key Laboratory of Quantum Technology and Device, Department of Physics, Zhejiang University, Hangzhou, China.
2. Department of Chemistry, Nagoya University, Nagoya 464-8602, Japan
3. Arts and Sciences, NYU Shanghai, Shanghai 200122, China
4. NYU-ECNU Institute of Physics at NYU Shanghai, Shanghai 200062, China
5. Department of Physics, Osaka University, Toyonaka 560-0043, Japan
6. Department of Physics, University of California at Berkeley, Berkeley, California 94720, USA.
7. Materials Science Division, Lawrence Berkeley National Laboratory, Berkeley, California, USA.
8. Kavli Energy NanoSciences Institute at University of California Berkeley and Lawrence Berkeley National Laboratory, Berkeley, California 94720, USA.

**Corresponding authors:**

Correspondence should be addressed to Dr. Sihan Zhao (sihanzhao88@zju.edu.cn), Dr. Ryo Kitaura (r.kitaura@nagoya-u.jp) and Prof. Feng Wang (fengwang76@berkeley.edu).

**#####All the authors contribute equally to the work.**

**Abstract:**

Different atomistic registry between the layers forming the inner and outer nanotubes can form one-dimensional (1D) van der Waals (vdW) moiré superlattices. Unlike the two-dimensional (2D) vdW moiré superlattices, effects of 1D vdW moiré superlattices on electronic and optical properties in 1D moiré superlattices are not well understood, and they are often neglected. In this Perspective, we summarize new experimental observations and theoretical perspectives related to interlayer interactions in double-walled carbon nanotubes (DWNTs), a representative 1D vdW moiré system. Our discussion will focus on new optical features emerging from the interlayer electronic interactions in DWNTs. Exciting correlated physics and exotic phases of matter are anticipated to exist in 1D vdW moiré superlattices, analogous with those discovered in the 2D vdW moiré superlattices. We further discuss the future directions in probing and uncovering interesting physical phenomena in 1D moiré superlattices.

# 1. Introduction

van der Waals (vdW) materials represent a large number of natural and artificial layered materials in which the stacking layers are weakly coupled by the vdW force, yet the atoms within each layer are ionically or covalently bonded.[1,2] Due to the weak bonds between the layers many vdW materials can be exfoliated down to atomically-thin layers.[3-7] Remarkably, the vdW interlayer interactions in layered materials with different stacking orders can lead to profound differences in electronic properties. For example, Bernal-stacked graphite[8-10] and rhombohedral graphite[11-13], two stable forms of crystalline graphite, exhibit very different electronic properties. When graphite is thinned down to the monolayer, the absence of interlayer interactions makes electrons in graphene behave as massless Dirac fermions,[14-16] in sharp contrast to the massive fermions in few layer graphene.[8,12] Significant effect of interlayer interactions was also found in other vdW layered materials such as the transition metal dichalcogenides (TMDCs),[5,17-19] transition metal halides,[20-22] metal phosphorus trichalcogenides etc.[21-23]

The successful isolation of two-dimensional (2D) layered materials together with the development of vdW assembly technique[24,25] have enabled broad exploration of novel physical phenomena in 2D vdW heterostructures. The vdW heterostructures do not have to satisfy the stringent lattice matching requirements. Moiré superlattices between layers have provided a powerful tool to engineer the electronic band structure and properties of the fabricated 2D vdW materials that are otherwise difficult.[26-39] Many fascinating new phases of matter have recently been discovered in 2D moiré superlattces,[32,33,40-64] including Mott and superconductivity ground states in magic-angle twisted bilayer graphene (MATBG)[32,33,40,41] and ABC trilayer graphene aligned with hBN,[43,44] and moiré excitons[45-48] and Wigner crystal states[49-52] in TMDC moiré superlattices. Interlayer coupling in moiré superlattices leads to the formation of moiré flat minibands.[29,53-59] Consequently, the on-site and/or long-range Coulomb interactions can become dominant over the kinetic energy (i.e., the bandwidth) and lead to strongly correlated electronic states. 2D vdW moiré superlattices also provide a unique playground for studying the non-trivial topological properties in a strongly correlated electronic system in which many-body electron-electron Coulomb interactions can induce symmetry-breakings.[60-64] Realization of an electrical control of robust magnetic states (pure orbital magnetism)[60-62] and superconducting switches in working devices[65-67] deliver promise of strongly-correlated 2D moiré superlattices in the application of electrically-writable magnetic memory and quantum information computing using tunable superconducting circuits. The works displayed in figure 1 represent a summary of recent experimental findings of moiré pattern-induced correlated physical phenomena in systems of MATBG (Figures. 1a-1c) and TMDC (Figures. 1d&1e) vdW heterostructures.

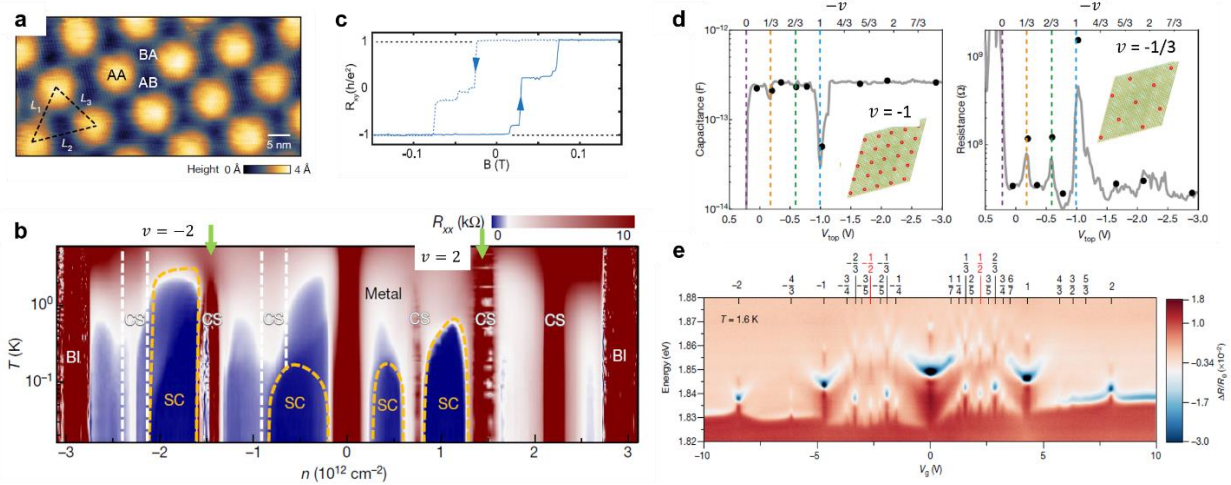

**Figure 1. Strongly correlated phenomena induced by interlayer coupling in 2D moiré superlattices. a)** Twisted bilayer graphehe near the magic angle (i.e. MATBG) has triangular lattice with alternating AA and AB/BA stacking orders in the moiré superlattice. AA sites appear higher in the STM topology and hold large local density of states (DOS). **b)** Phase diagram of MATBG. Correlated states (CS) appear at all the integer filling of the moiré band, which are accompanied by superconducting (SC) phases. Two green arrows indicate the integer filling at $v = \pm 2$. **c)** Quantum anomalous Hall (QAH) showing quantized Hall resistance in MATBG aligned with hBN at an integer filling $v = 3$. The observed quantized Hall signals at B = 0 comes from a pure orbital magnetism and indicates an incipient topological Chern insulator with interaction-induced time-reversal symmetry breaking. **d)** Mott insulating states ($v = -1$ half-filling the moiré hole band) and generalized Wignal crystal insulating states (fractional filling at $v = -1/3$ and $v = -2/3$) were observed in angle-aligned TMDC ($WS_2/WSe_2$) moiré superlattices by an effective capacitance measuremnt. Insets shows the localized hole position (red balls) on the moiré superlattice at $v = -1$ and $v = -1/3$. **e)** A diverse set of Wigner crystal states beyond 1/3 and 2/3 fillings also exist in the $WS_2/WSe_2$ moiré system. a) Adapted with permission.[42] Copyright 2019, Springer Nature Limited. b) Adapted with permission.[41] Copyright 2019, Springer Nature Limited. c) Reproduced with permission.[61] Copyright 2020, American Association for the Advancement of Science. d) Adapted with permission.[49] Copyright 2020, Springer Nature Limited. e) Reproduced with permission.[51] Copyright 2020, Springer Nature Limited.

Compared with the exciting developments in 2D vdW moiré heterostructures, 1D moiré superlattices have been little explored. In the following, we first introduce 1D moiré superlattices and the interlayer interactions using a representative vdW material in 1D, double-walled carbon nanotubes (DWNTs). Then we briefly review previous results on mechanical and transport properties of structure-identified DWNTs. We then focus on optical spectroscopy studies of interlayer interactions in structure-identified DWNTs, which revealed new insights of the interlayer interactions in 1D moiré systems. Finally, we will discuss the future directions and outlook of studying the 1D moiré superlattices and 1D moiré physics.

## 2. Interlayer interactions in 1D moiré superlattices

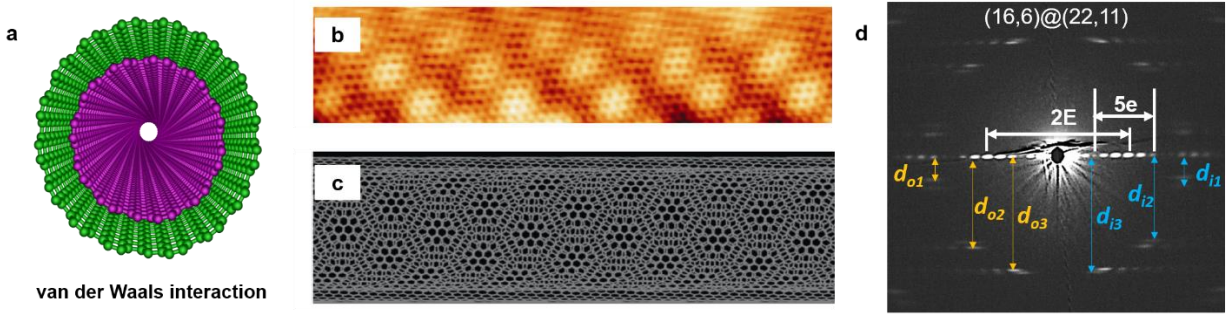

**Figure 2. van der Waals-coupled 1D DWNT moiré superlattice. a)** Structure model of a DWNT composed of two van der Waals-coupled SWNTs. **b)** STM topology image of an isolated DWNT showing spatially-modulated moiré pattern. **c)** The rigid lattice model of a DWNT reproducing the observed moiré pattern in b). SWNTs (25,20) and (21,32) are used in panel c for the inner and outer nanotubes in order to produce b). **d)** Electron diffraction pattern of an DWNT with chirality (16,6)@(22,11). There are total 4 sets of hexagons in its diffraction pattern, two each belonging to each nanotube. The equatorial line oscillates in intensity with E and e representing the long and short periods. Distances for features outside equatorial line to the equatorial line (e.g., $d_{o2}$ and $d_{o3}$ for outer SWNT and $d_{i2}$ and $d_{i3}$ for inner SWNT) are used to uniquely determine the DWNT chirality, together with the measured E and e. b, c) Reproduced with permission.[85] Copyright 2013, ELSEVIER. d) Adapted with permission.[89] Copyright 2014, Springer.

When two or more vdW layered materials are concentrically rolled up, it forms 1D moiré superlattices. Double-walled and multi-walled carbon,[68-72] boron nitride,[73,74] TMDC[75] nanotubes and their heterostructures[76,77] are among the available material list of 1D vdW moiré superlattices.

### 2.1. A model system of 1D moiré superlattices: DWNTs

Double-walled carbon nanotubes (DWNTs) provide one of the ideal model systems to study the interlayer interactions in 1D moiré superlattices because the physical properties of a DWNT depend on the combination of the two components,[78-81] non-interacting single-walled carbon nanotubes (SWNTs). The electronic property of each SWNT is physically determined by two chiral vectors. Conventionally, the two chiral vectors for inner and outer SWNTs are represented by $\boldsymbol{C}$ and $\boldsymbol{C'}$ (also known as chirality),[78] respectively, where

$$\boldsymbol{C} = n_1\boldsymbol{a_1} + n_2\boldsymbol{a_2} = (n_1, n_2) \text{ and } \boldsymbol{C'} = n_1'\boldsymbol{a_1} + n_2'\boldsymbol{a_2} = (n_1', n_2')  \quad (1),$$

with $a_1$ and $a_2$ denoting the unit lattice vectors of 2D graphene sheet. The chiral angle of a specific SWNT is defined as

$$\theta = \cos^{-1} \frac{2n_1 + n_2}{2\sqrt{n_1^2 + n_2^2 + n_1 n_2}} \quad (2),$$

constrained within $0° < \theta < 30°$. In DWNTs, the inner and outer SWNTs are coupled by vdW interactions (Figure 2a) with interlayer distance ranging from 0.30 – 0.40 nm.[82,83] It is still debatable if there is any correlation of chiral angles between the inner and outer SWNTs,[82-84] and the answer may depend on the synthesis method by which the DWNT samples are produced. The DWNT forms a generic 1D moiré superlattice in analogy with the 2D counterpart. Formation of 1D moiré superlattices has been clearly observed in real space by scanning tunneling microscopy (STM) (Figures 2b&2c).[85]

The physical structure (i.e., moiré superlattice) of an individual DWNT can only be unequivocally determined by electron beam diffractions.[82-84,86-89] Figure 2d shows a typical diffraction pattern of an isolated DWNT and how the chirality of two constituent SWNTs are determined[89]. The diffraction pattern of a typical DWNT exhibits two sets of mutually twisted hexagonal patterns belonging to two constituent SWNTs.[90-92] The equatorial line in the DWNT diffraction pattern (perpendicular to the DWNT axis in real space) exhibits a "beating-like" behavior with the fast oscillations (with period e) modulated by slow ones (with period E). This originates from the interference of electron waves scattered by two nanotubes in the radial direction. Because of the high sensitivity of the interference to diameter and/or chirality of each nanotube, the equatorial line profile serves as a decisive feature for the unambiguous determination of DWNT structure. By measuring the distances from features outside the equatorial line to the equatorial line, for example, $d_{o2}$ and $d_{o3}$ from outer SWNT and $d_{i2}$ and $d_{i3}$ from inner SWNT, the DWNT physical structures can be fully obtained by using a set of equations:

$$n_1 = \frac{\pi}{\sqrt{3}}(2d_{i3} - d_{i2})(\frac{1}{e} - \frac{1}{E}), \; n_2 = \frac{\pi}{\sqrt{3}}(2d_{i2} - d_{i3})(\frac{1}{e} - \frac{1}{E}), \; n_1' = \frac{\pi}{\sqrt{3}}(2d_{o3} - d_{o2})(\frac{1}{e} + \frac{1}{E}), \; n_2' = \frac{\pi}{\sqrt{3}}(2d_{o2} - d_{o3})(\frac{1}{e} + \frac{1}{E}) \; (3).^{[88]}$$

The determined structure of Figure 2d using Eq. (3) is (16,6)@(22,11),[89] whose result is further compared and corroborated by simulations.

Earlier theoretical studies have focused on commensurate DWNTs or SWNT bundles comprising armchair or zigzag SWNTs, showing significant changes of the band structure from interlayer interactions.[93-96] However, such commensurate DWNTs have been very rarely realized experimentally.[84] For the commonly observed incommensurate DWNTs conventional wisdom suggests that the interlayer interaction is very weak. This is because the coupling matrix element is averaged out to zero when summing up the oscillating interlayer hopping at each lattice site.[97,98]

1D materials such as SWNTs show characteristic paired subbands associated with van Hove singularity (VHS) of density of states (DOS).[78] This unique property makes the optical spectroscopy tool very powerful in probing the interlayer interactions in 1D systems. In fact, early photoluminescence studies consistently revealed energy redshifts of optical transition energies for the inner semiconducting in DWNTs[99-102] and SWNTs embedded in a dielectric environment[103,104]. The environmental screening effect is accounted for the observed energy redshifts.[105-107] These results support the weak coupling in incommensurate DWNTs where one SWNT wall behaves no more than a dielectric screening layer of the other SWNT.

Before focusing on the new optical spectroscopy discoveries in interlayer electronic interactions in DWNTs, we first briefly review the works on the interlayer mechanical coupling as well as on the transport studies in the structure-identified DWNTs, showing signs of non-trivial interlayer coupling.

## 2.2. Mechanical coupling in DWNTs

Raman spectroscopy serves as a valuable tool in studying the vibrational properties and characterizing graphitic materials.[108,109] It is widely used to characterize the diameters and chirality information of SWNTs,[78] DWNTs,[110] and MWNTs[111] in which the observed radial breathing modes (RBM) can be directly connected with the diameter of each nanotube by $\omega_{RBM} = 228/d$, where $\omega_{RBM}$ is in cm$^{-1}$ and $d$ is in nm.[78] However, the use of such relation was later found problematic when carefully examining the individual DWNTs with known chirality.[112,113] Figures 3a-3c investigated Raman RBM modes on a (12,8)@(16,14) DWNT with varying laser excitation energies.[112] The results shown in Figures 3d-3e examined different chirality-identified DWNTs with a single laser line excitation.[113] Both reported the RBM positions in DWNTs (denoted by $\omega_L$ and $\omega_H$) are very different from those expected from isolated inner and outer SWNTs of the same chirality ($\omega_i$ and $\omega_o$ indicated by the black dashed lines). These two studies also found that

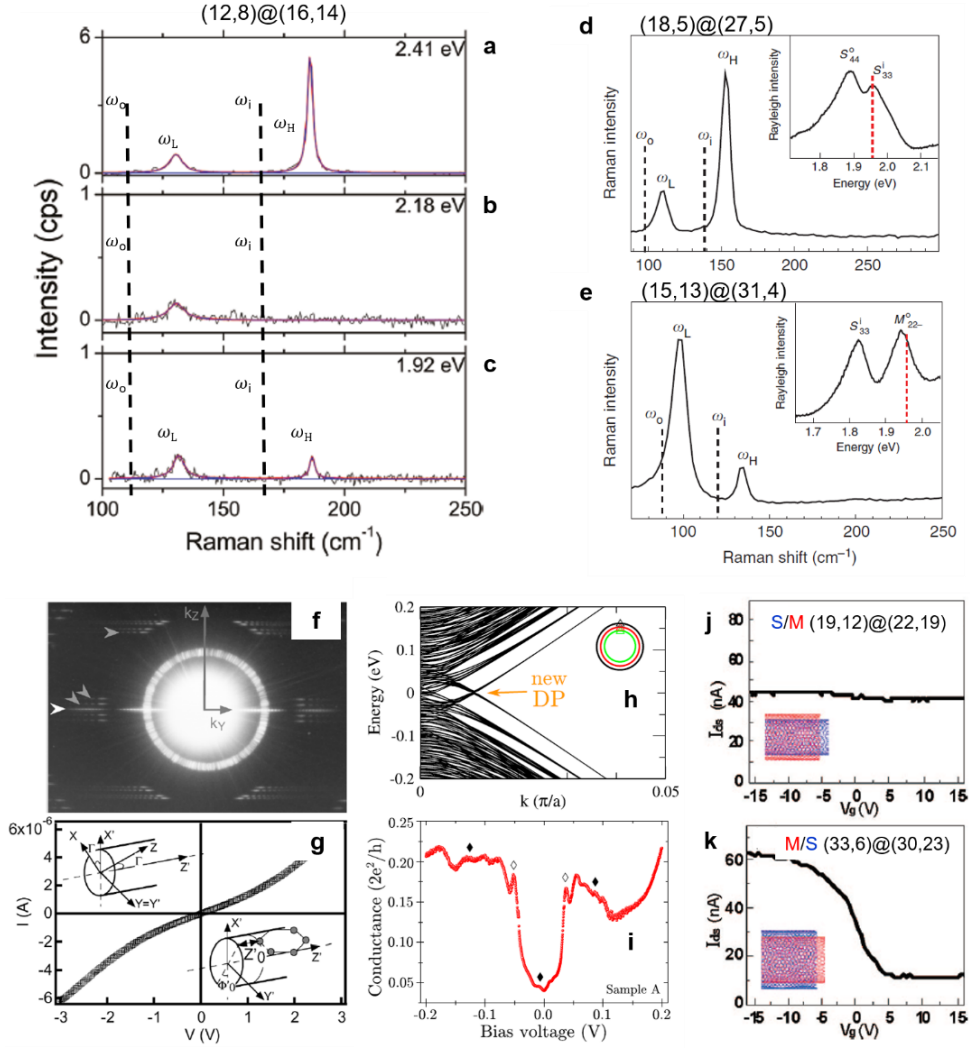

**Figure 3. Mechanical and electrical transport properties of structure-identified DWNTs. a, b, c)** Raman RBM modes for a (12,8)@(16,14) DWNT with laser excitation at 2.41 eV, 2.18 eV and 1.92 eV, repsectively. Excitations at 2.41 eV, 2.18 eV and 1.92 eV are in close vicinity to $S_{33}$ of inner (12,8), $S_{44}$ of outer (16,14) and $S_{33}$ of outer (16,14), respectively. The observed coupled modes $\omega_L$ and $\omega_H$ are very different from those of uncoupled SWNTs ($\omega_i$ and $\omega_o$ indicated by the dashed lines). **d, e)** Raman RBM modes for (18,5)@(27,5) and (15,13)@(31,4) DWNTs, respectively, with a single laser excitation at 1.96 eV (red dashed line in the inset). **f)** Electron beam diffraction pattern of a semiconducting DWNT (51,34)@(53,43). **g)** I-V curve of the same DWNT in f) measured inside a transmission electron microscope (TEM) in ultrahigh vacuum. Linear I-V near zero bias indicates a finite DOS near the Fermi surface. **h)** Tight-binding band structure of a 80 nm diameter zigzag TWNT (1045, 0)@(1054, 0)@(1063, 0), where formation of a new Dirac point (DP) is indicated by an arrow. **i)** Measured tunneling conductance as a function of bias voltage between an isolated TWNT and metal contact at 4.2 K. The observed conductance singularities (white diamonds) and oscillations (black diamonds) are assigned to opening of new Dirac points (DP) caused by the formed 1D moiré superlattice. **j, k)** Gate-dependent current of two structure-identified DWNTs (19,12)@(22,19) (j) and (33,6)@(30,23) (k). The effect of interlayer electronic interactions is not noticeable. a-c) Adapted with permission.[112] Copyright 2011, American Chemical Society. d, e) Adapted with permission.[113] Copyright 2013, Springer Nature Limited. f, g) Reproduced with permission.[117] Copyright 2002, American Physical Society. h ,i) Reproduced with permission.[118] Copyright 2016, Springer Nature Limited. j, k) Adapted with permission.[119] Copyright 2009, American Chemical Society.

$\omega_L$ and $\omega_H$ appeared whenever the excitation hits on the optical resonance in either of two

nanotubes (Figure 3b, however, does not hold). The coupled modes (i.e., in phase and out of phase modes) can be theoretically described by two coupled mechanical oscillators.[113,114] These observations show clear evidences of a strong mechanical coupling between vdW-coupled SWNTs, making people aware of DWNT distinct vibrational properties than simply adding up those of two-component SWNTs. Similar effect of the mechanical coupling was also reported for G modes at higher wavenumbers.[115] Evidence of moiré-induced vibrational coupling in DWNTs was recently reported,[116] highlighting the importance of 1D moiré superlattices.

### 2.3. Electrical transport studies in DWNTs

Transport measurement is suitable to probe the electronic band structure at the Fermi energy. The first experimental evidence showing the possible change of DWNT band structure was reported in 2002 by an in-situ transport study (Figures 3f and 3g).[117] The authors observed very puzzling finite DOS at Fermi surface (zero bias region in the I-V curve shown in Figure 3g), which normally should be gapped for a DWNT (51,34)@(53,43) consisting of both semiconducting inner and outer SWNTs (Figure 3f). It indicated the possibility of notable band structure change induced by the interlayer electronic interactions, although the exact mechanism was not understood. A more recent transport (tunneling) study (Figures 2h&2i) reported extra conductance singularities and oscillations in an isolated MWNT at 4.2 K (indicated by white and black diamonds in FIG. 2i).[118] Band structure shown in Figure 2h was calculated using a hypothesized MWNT structure. The authors argued the moiré superlattice-induced opening of new Dirac points (DP) (Figure 2h) accounts for the appearance of these additional conductance in Figure 2i. This is a very interesting result, but the lack of definitive structure information and the complicated contact between the MWNT and metal leads make a definitive physical interpretation difficult. There were other transport studies on structure-identified individual DWNTs prepared by either a direct growth[119] or sorted technique[120,121]. Typical results of gate-dependent conductance are shown in Figures 3j and 3k.[119] The observed transfer characteristics can be mostly described by outer shell-contacted DWNTs in S@M (Figure 3j) and M@S configurations (Figure 3k) without considering interlayer electronic coupling.[119-122] In summary, transport measurements reported so far have not found conclusive evidence of effect of 1D moiré superlattices in DWNTs.

### 3. Optical spectroscopy studies on interlayer interactions in DWNTs
### 3.1. Interactions in the perturbative regime

Progress continued after successful applications of broadband optical spectroscopy technique (either Rayleigh scattering spectroscopy or absorption spectroscopy) on individual SWNTs and DWNTs suspended in air.[123-130] Typical samples are directly grown by chemical vapor deposition method across a long trench on $SiO_2$/Si substrates with a careful density control.[131,132] Figure 4a shows the schematic of Rayleigh scattering spectroscopy technique used to probe the optical resonances of nanotubes. Broadband light from a supercontinuum laser source (e.g., 1.2–2.7 eV) is focused on the suspended nanotube samples with laser polarized along the nanotube axis. The light scattered by the nanotube is collected by a second objective and directed to a CCD camera for imaging and a spectrograph for spectroscopy measurement. The optical transition energies coming from paired 1D subbands (denoted by $S_{11}$, $S_{22}$, $M_{11}$, $S_{33}$ etc.) in each SWNT can be precisely determined by using the optical spectroscopy technique. The combined broadband spectroscopy and electron microscopy (determine the chirality) on individual nanotube level (schematic drawing shown in Figure 4b) enables a clear observation of the effect of interlayer electronic couplings when comparing to the established optical spectra of corresponding SWNTs of the same chirality.[81]

Figures 4c-4g show the results from [98] of optical resonances in chirality-defined DWNTs. The experimentally measured optical absorption of an DWNT is shown in Figure 4d whose structure has been identified as (11,11)@(22,9) (diffraction pattern shown in Figure 4c). Four absorption peaks in Figure 4d are identified: three peaks come from the optical transitions ($S_{33}$, $S_{44}$, $S_{55}$) of outer semiconducting nanotube with chirality (22,9) and the other peak ($M_{11}$) comes from the inner metallic nanotube with chirality (11,11). All the peaks exhibit energy redshifts when comparing with corresponding SWNTs of the same chirality (dashed lines). The statistical result of energy shift for 99 optical transitions is presented in Figure 4e (solid circles). It is clear that both energy red and blue shifts (- 200 meV to + 50 meV) exist, being consistent with other reports.[89]

The observed energy blue shift was not compatible with the dielectric screening effect used to explain previous reported energy redshifts.[99-107] The authors further proposed the first theoretical model to explain and simulate the observed energy shifts in 1D incommensurate DWNTs.[98] The model is based on a calculation of electronic coupling matrix element $M_{\alpha\beta} = \langle \Psi_\alpha | H_{INT} | \Psi_\beta \rangle$ between the tight-binding electronic wavefunctions of inner ($\Psi_\alpha$) and outer ($\Psi_\beta$) nanotubes. Shown in Figure 4f, the extended 2D graphene Brillouin zone (GBZ) of outer layer (black) is rotated and uniaxially stretched (compressed in real space) along nanotube circumferential direction ($C$ and $C'$ direction in Figure 5b) relative to the unchanged GBZ of the inner layer (pink). In this case, electronic states (in k space) from inner ($q_\alpha$) and outer SWNT ($q_\beta$) can have electronic coupling, that is, the states on cutting lines from two nanotubes match exactly (see right panels in Figures 5g&5i). With the knowledge of $M_{\alpha\beta}$ and by using the perturbation theory (to the second order), the energy shifts of, for example, inner optical transitions $\Delta E_{ii}$ caused by the outer layer can be calculated as:

$$\Delta E_{ii} = \sum_{\beta=1}^{3} \frac{|M_{\alpha\beta}|^2}{E_\alpha - E_\beta} \quad (4),$$

where $E_\alpha$ and $E_\beta$ are eigen energies of electronic states $q_\alpha$ (inner) and $q_\beta$ (outer) with $q_\beta = q_\alpha + G_i$ and $G_i$ the reciprocal lattice vector of pristine GBZ. Note that the sum in Eq. (4) is taken over three states of outer nanotube that is closest to Γ point (three red circles in Figure 4f) as an approximation because of the small contributions from other states (black circles in Figure 4f) abiding to the exponential decay of coupling strength. This theory can reproduce the experimental results reasonably well as shown in Figure 4e. As an example, Figure 4g presents the simulation of the optical transitions ($S_{33}$ and $S_{44}$) for a (15,10) inner SWNT with varying the chirality of outer SWNT. It shows that the amplitude of energy shift depends sensitively on the specific optical transitions and the outer tube species, which can be either positive or negative with a magnitude as large as 150 meV.

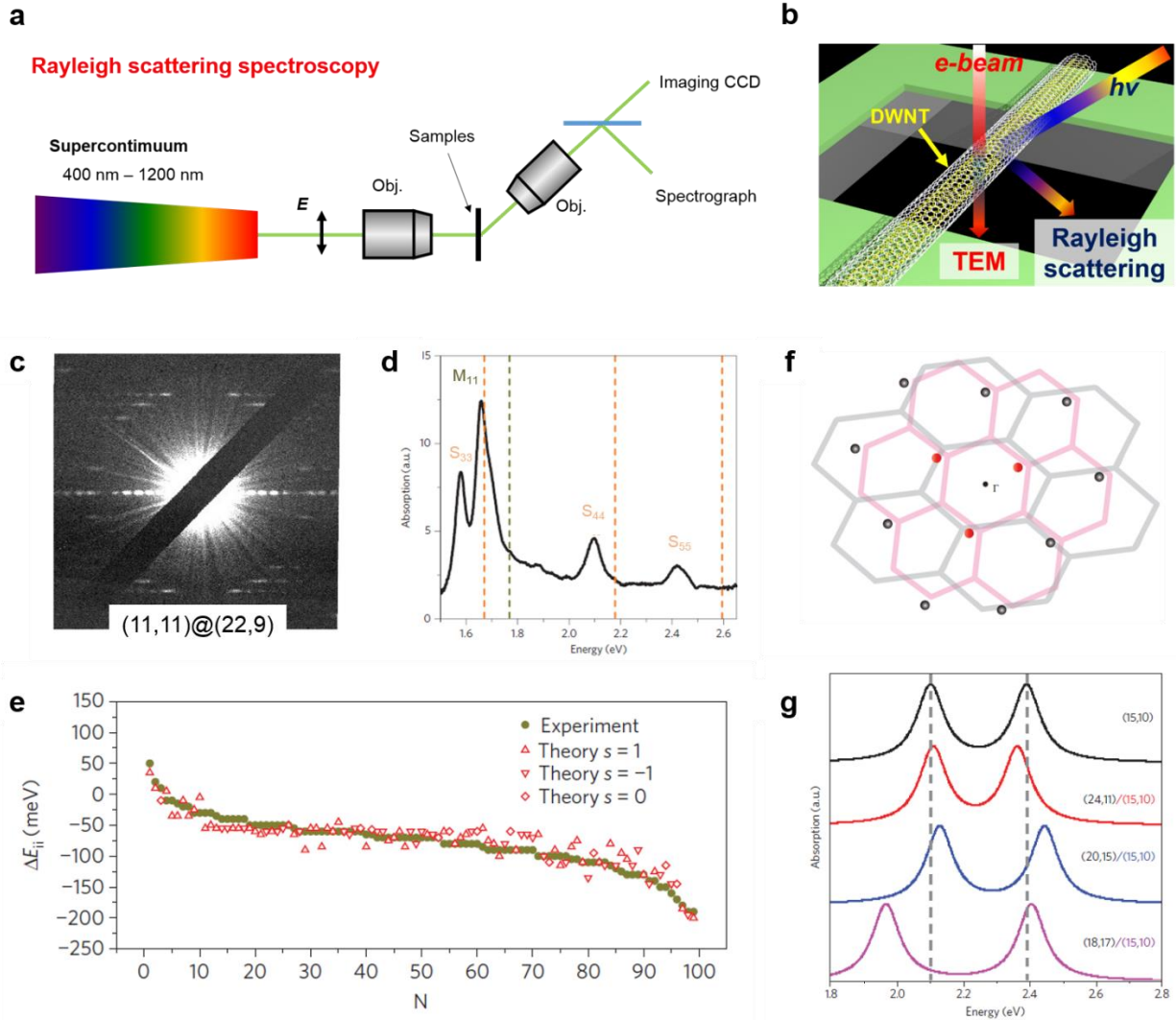

**Figure 4. Interlayer electronic interactions probed by optical spectroscopy for structure-identified incommensurate DWNTs. a)** Schematic of Rayleigh scattering spectroscopy setup used to probe optical transitions of nanotubes. **b)** Experimental scheme showing the combined study of optical spectroscopy and electron beam diffraction on individual DWNTs. **c)** Electron beam diffraction pattern of an DWNT with chirality (11,11)@(22,9). **d)** Absorption spectrum of the same DWNT shown in panel c. Three optical transitions ($S_{33}$, $S_{44}$ and $S_{55}$) come from outer semiconducting SWNT (22,9) and one ($M_{11}$) comes from inner metallic SWNT (11,11). Comparing to the optical transitions of isolated SWNTs (22,9) and (11,11) (indicated by the dashed lines), all the four optical transitions in DWNT exhibit energy redshifts. **e)** Summary of energy shifts for 99 optical transitions in different DWNTs. Experimental results are presented by green solid dots, and the predicted values using the developed perturbation theory are shown by red markers. **f)** Schematic drawing of extended 2D GBZ depicting the physics of interlayer electronic interactions between inner (in pink) and outer (in black) nanotubes. Here GBZ of outer layer (in black) is uniaxial distorted to satisfy the electronic coupling conditions (see the text). Important electronic states included in the perturbation theory (Eq. (4)) are indicated by the red dots which are the three closest states around Γ point; contribution from other possible state is small and indicated by the black dots. **g)** Simulation of the interlayer electronic interaction-induced optical transitions ($S_{33}$ and $S_{44}$) for a (15,10) inner SWNT with varying the chirality of outer SWNT using Eq. (4). b) Adapted with permission.[89] Copyright 2014, Springer. a-h) Adapted and reproduced with permission.[98] Copyright 2014, Springer Nature Limited.

### 3.2. Interactions in the non-perturbative regime

The interlayer interactions in DWNT moiré superlattices discussed above are still limited within a perturbative regime. On the other hand, a recent work,[134] motivated by [133], theorized and calculated the band structure of certain species of DWNTs with moiré superlattice effect beyond the perturbative coupling regime. The schematics and lattice structure presented in Figures 5a-5f illustrate how a 1D DWNT moiré superlattice (Figures 5c&5f) is constructed from two AA-stacked graphene nanoribbons (Figures 5a&5d). First, align two chiral vectors $C$ and $C'$ by rotating the outer ribbon relative to the inner one (Figures 5b&5e) in which the rotation angle is determined by the chiral angle difference between two SWNTs (Eq. (2)). It is followed by a uniaxial compression of the outer ribbon such that $C'$ becomes equal to $C$ in length (Figures 5c&5f). We note that the formed 1D moiré superlattice (Figure 5f) is distinct from 2D moiré pattern of TBG (Figure 5e) due to the unique uniaxial compression in 1D case.[98,134-137] The chiral vectors of the inner and outer tubes have different lengths, but their registry along the circumferential direction should be commensurate. This makes the lattice structure "virtually" strained along that direction (Figure 5f), and enables some interaction which is not allowed in the stack of two hexagonal lattices in 2D. In k space, it results in a rotated and uniaxial stretched GBZ for the outer layer with an intact GBZ for the inner layer (also see Figure 4f). We also note that the intertube distance of a DWNT can vary over a wider range between 0.30 nm – 0.40 nm[82,83] depending on the combination of chiralities, while the interlayer distance of TBG only slightly varies with the twist angle.[138] The tubes with shorter intertube distance will have much stronger interaction than those with longer distance.

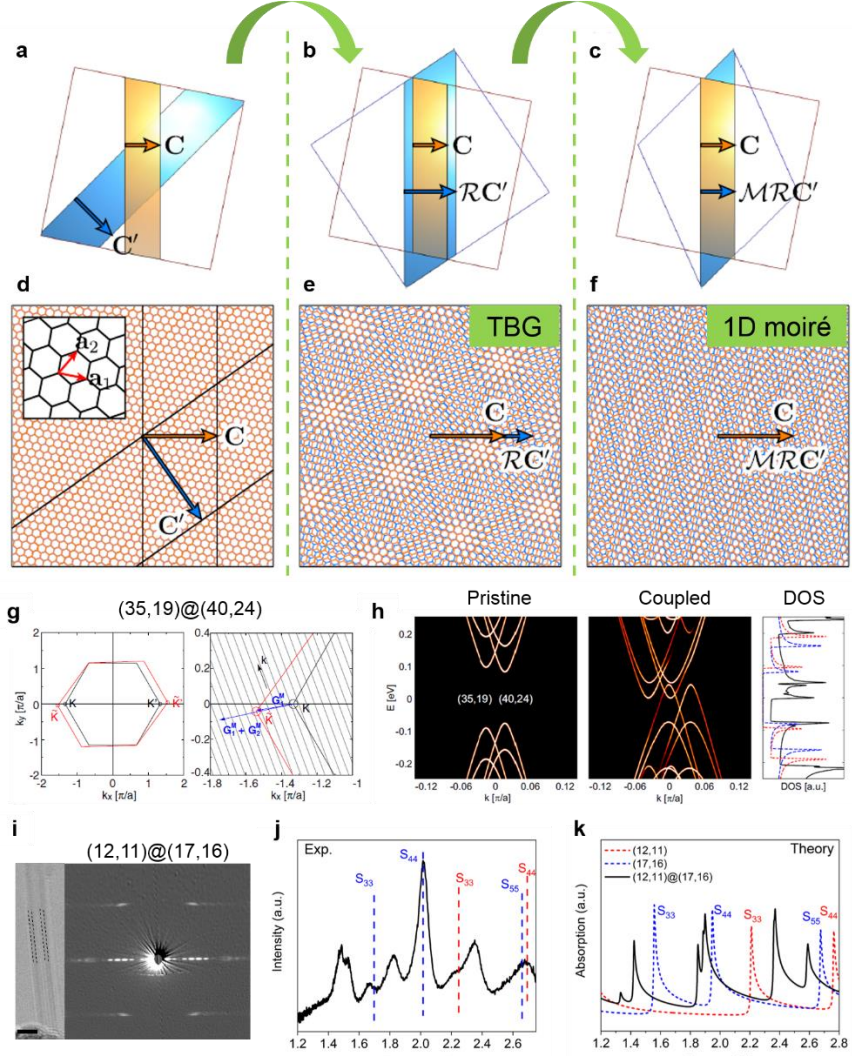

**Figure 5. Theoretical model and experimental observation of non-perturbative electronic interactions in 1D DWNT moiré superlattices. a, d)** Schematic drawing and lattice structure of two AA stacked graphene nanoribbons. **b, e)** Schematic drawing and 2D moiré superlattice structure after a rotation operation of outer ribbon ($\mathcal{R}C'$) such that $C$ and $C'$ align. **c, f)** Schematic drawing and 1D moiré superlattice structure after a consecutive uniaxial compression of outer ribbon ($\mathcal{MR}C'$). The inner ribbon is presented in orange (orange lattice) with $C$ denoting the chiral vector of inner SWNT, while outer ribbon is shown in blue (blue lattice) with $C'$ denoting the chiral vector of outer SWNT. **g)** 2D GBZ and its enlarged view of DWNT (35,19)@(40,24) that satisfies the strong-coupling condition exactly ($\Delta K = G_1^M$). **h)** Calculated band structures of two pristine SWNTs (left panel) and coupled DWNT (35,19)@(40,24) (middle panel). The corresponding DOS plot is shown in the right panel with dashed lines denoting DOS of two pristine SWNTs and black line denoting that of the coupled DWNT. **i)** TEM image and diffraction pattern of a DWNT satisfying the predicted strong-coupling condition. The structure of the DWNT is determined as (12,11)@(17,16) by analyzing the diffraction pattern and further corroborated with simulations. The scale bar in TEM image is about 2 nm. **j)** Measured Rayleigh scattering spectrum of DWNT (12,11)@(17,16). Optical transitions from pristine SWNT (17,16) and inner (12,11) are indicated by the blue and red dashed lines, respectively. The optical spectrum is very different from sum of two constituent SWNTs, indicative of moiré superlattice-induced strong electronic coupling. **k)** Theoretically calculated optical absorption spectrum of (12,11)@(17,16) with (solid black line) and without (blue and red dashed lines) the moiré effect within the proposed theoretical framework[134]. Note that the calculated optical absorption peaks of pristine SWNTs very well reproduced the experiments within an accuracy of ~ 100 meV (Figures. 5b&5c). a-h) Adapted and reproduced with permission.[134] Copyright 2015, American Physical Society. i-k) Adapted and reproduced with permission.[139] Copyright 2020, American Physical Society.

Within the theoretical framework proposed in [134], the interlayer electronic coupling in DWNT moiré superlattices is fully characterized by the relative orientation between $\boldsymbol{C}$ and $\boldsymbol{C}'$. The reciprocal lattice vectors in DWNT moiré superlattice have the following form

$$\boldsymbol{G}_i^M \cdot \boldsymbol{C} = 2\pi(n_i - n_i')(i = 1,2) \quad (5),$$

where $\boldsymbol{C}$, $n_i$ and $n_i'$ are defined in Eq. (1). In FIG. 5g the 2D GBZ of the outer layer is presented by distorted red hexagons (Dirac points $\widetilde{K}$ and $\widetilde{K}'$) while that of the inner layer is shown by the undistorted black hexagons (Dirac points $K$ and $K'$). The states of nanotubes are quantized, resulting in the cutting lines shown in Figure 5g. Significant modification of the band structure can take place under two different conditions, which are dubbed the strong-coupling case and the flat-band case, respectively. The former occurs when $\boldsymbol{C}$ and $\boldsymbol{C}'$ are nearly parallel to each other and, at the same time, the difference of two chiral vectors ($\boldsymbol{C}' - \boldsymbol{C}$) is parallel to the armchair direction. Then the moiré superlattice potential makes the resonant coupling between the states of constituent nanotubes, leading to a drastic energy shift and splitting of the subband edges. The latter case occurs under the condition that $\boldsymbol{C}$ and $\boldsymbol{C}'$ are nearly parallel, and $\boldsymbol{C}' - \boldsymbol{C}$ is parallel to the zigzag direction. A long-period moiré interference potential turns the original single nanotube bands into a series of nearly flat bands, reminiscent of moiré flat bands in TBG at small twist angle.[29,134] The presence of strong-coupling and flat-band cases is a direct consequence of the strained structure along the circumferential direction that is unique to 1D superlattices.

We present below a simple derivation of the strong-coupling case. The interacting Hamiltonian for low-energy electrons can be written as

$$H = \begin{pmatrix} H_1(k) & U^\dagger \\ U & H_2(k) \end{pmatrix} \quad (6)$$

, where situation around one valley (e.g., $K$ valley) is considered. $U$ is interlayer coupling matrix element with the form $\langle k', X'_{l'} | T | k, X_l \rangle$, where $|k, X_l\rangle$ is an intralayer Bloch wave basis, $X$ and $X'$ are either A or B atom in graphene, $l$ and $l'$ are either 1 or 2 (inner or outer layer), and $T$ is the interlayer coupling Hamiltonian. If the formed moiré period is much longer than the graphene lattice (continuum model), the interlayer electronic coupling $U$ can be simplified consisting of three Fourier components of 1, $e^{i\boldsymbol{G}_1^M \cdot \boldsymbol{r}}$ and $e^{i(\boldsymbol{G}_1^M + \boldsymbol{G}_2^M) \cdot \boldsymbol{r}}$. The coupling effect will be significant when the distance between $K$ and $\widetilde{K}$ points, $\Delta \boldsymbol{K} \equiv \widetilde{\boldsymbol{K}} - \boldsymbol{K} = (2\boldsymbol{G}_1^M + \boldsymbol{G}_2^M)/3$, is close to any of the three Fourier components of 0, $\boldsymbol{G}_1^M$ and $\boldsymbol{G}_1^M + \boldsymbol{G}_2^M$. This mechanism leads to the criteria showing the strong-coupling effect, which was recently observed in an experiment (discuss later). A semiconducting DWNT with chirality (35,19)@(40,24) matches the predicted condition of strong-coupling, i.e., $\Delta \boldsymbol{K} = \boldsymbol{G}_1^M$ (Figure 5g). Its band structure (Figure 5h) shows a drastic change: the original gap vanishes, exhibiting a finite DOS at the Fermi energy and that extra number of VHS in the DOS plot shown on the right.

A very recent experiment reported the drastic change of electronic structure in a DWNT moiré superlattice.[139] The experiment was carried out by a combined Rayleigh scattering spectroscopy and electron beam diffraction on individual ultraclean DWNTs suspended in air (experimental scheme shown in Figure 4b). The Rayleigh scattering spectra are obtained by normalizing the measured scattered light with the incident laser profile. Figure 5i shows the TEM image and the diffraction pattern of this special DWNT. The chiral index is unequivocally identified as (12,11)@(17,16) from the diffraction pattern, which is further corroborated by systematic simulation. The determined structure matches the theoretical criteria of strong-coupling exactly since $\boldsymbol{C}$ and $\boldsymbol{C}'$ are both nearly armchair with a chiral angle difference of about 0.4° and that $\boldsymbol{C}' - \boldsymbol{C} = (5,5)$ orients along the armchair direction. Eight well-defined optical resonances were observed within energy 1.35-2.7 eV in its optical transition spectrum (Figure 5j). The expected optical transition energies of two pristine SWNTs comprising the DWNT are indicated by the dashed blue (outer SWNT) and red (inner SWNT) lines.[81] At a glimpse, the optical spectrum is very puzzling because the peaks are completely different from the optical transitions of two constituent SWNTs. Several additional optical resonances emerge in the DWNT optical spectrum, which definitely cannot be explained by a perturbation theory (e.g., Eq. (4)). Using the theoretical framework including the non-perturbative moiré superlattice effects,[134] the calculated optical absorption spectra of (12,11)@(17,16) is shown as the solid black line in Figure 5k. The theoretically calculated optical spectrum (solid black line in Figure 5k) qualitatively reproduces the experiment (Figure 5j). The calculated absorption of pristine SWNTs (12,11) and (17,16) are also shown by the dashed red and blue lines, respectively. Note that the calculated optical absorption peaks of pristine SWNTs very well reproduced the experiments within an accuracy of ~ 100 meV (Figures 5j&5k). The theory shows that the moiré superlattice-induced interlayer coupling strongly mixes the electronic states in two SWNTs, resulting in a very different band structure and thus a drastic change of the optical spectrum.

Several theoretical reports quickly followed the above observation of 1D moiré superlattice effect.[140-144] One study predicted strong inter-tube optical transitions in DWNT moiré superlattices when satisfying certain structure requirements.[140] It claimed that the interlayer electronic coupling can modify the band structure in quite a few DWNTs and shift the VHS positions of each constituent SWNTs (Figure 6a), making inter-tube transitions optically bright and dominate over the intra-tube transitions (indicated by the arrows in the right panel of Figure 6a). It is now well accepted that new optical features can emerge in DWNTs due to strong coupling of interlayer electronic states.

## 4. Future outlook and summary

There are several important questions to be answered in the study of 1D moiré superlattices.

### 4.1. Moiré excitons

Does well-defined moiré excitons exist in DWNTs and other 1D superlattices, and are they spatially localized? The normal band alignment in a DWNT is type I, that is to say, the conduction and valence band of outer SWNT energetically lie within those of inner SWNT. Therefore, in principle, only intra-layer excitons (i.e., electron and hole reside in the same SWNT) can exist if we neglect the band structure reconstruction. The intralayer excitons can be preferentially trapped by the moiré superlattice potential at certain regions inside the moiré unit cell where a global or local energy minimum is located.[56,145,146] Exciton localization due to the moiré effect has been observed in 2D TMDC moiré superlattices.[147,148] The realization of 1D moiré excitons can potentially realize the attractive 1D array of zero-dimensional quantum emitters. On the other hand,

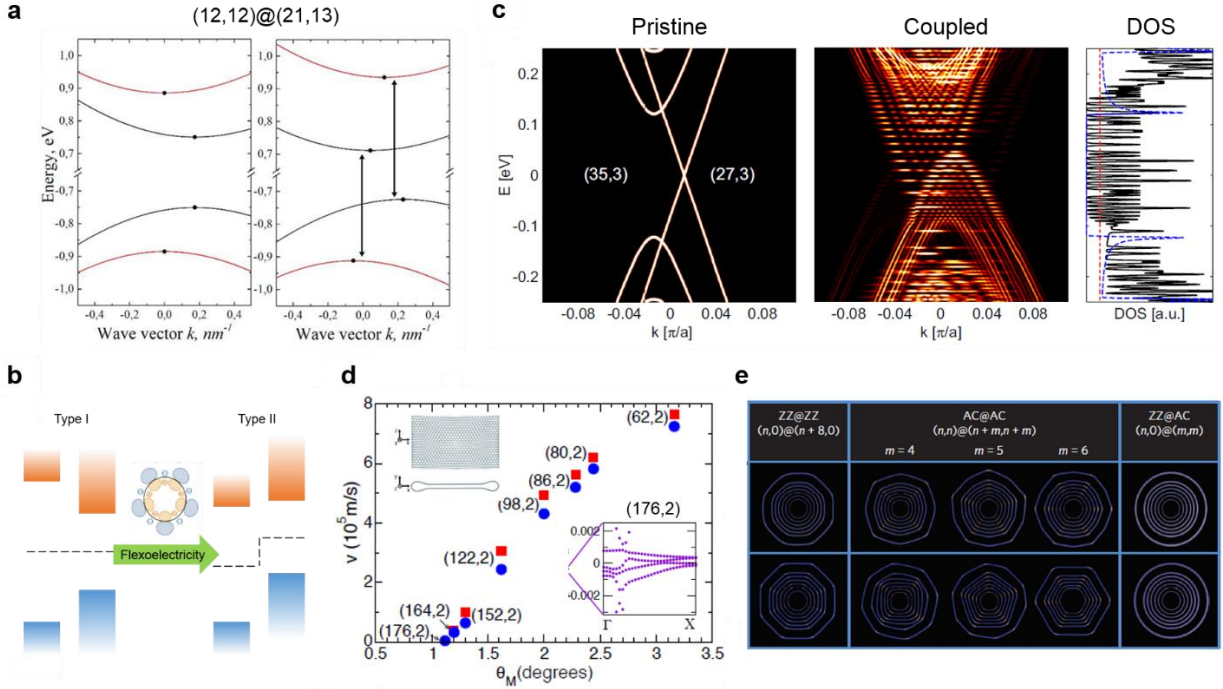

**Figure 6. Theoretical prediction and understanding of physical phenomena related to 1D moiré superlattices. a)** Dispersion relation of a DWNT (12,12)@(21,13) without (left panel) and with (right panel) effect of moiré superlattice. Dispersion curves of inner and outer nanotubes are shown in red and black, respectively. VHS positions are indicated by black dots in which VHS of pristine outer SWNT is taken as the origin ($k = 0$). The interlayer coupling shifts the VHS positions and modifies the DWNT optical spectra, which leads to the inter-tube optical transitions (indicated by arrows). **b)** Flexoelectricity-induced change of band alignment from type II to type I predicted for DWNTs with diameter larger than 2.4 nm. **c)** Formation of flat bands in vdW-coupled DWNT with chirality (27,3)@(35,3). **d)** Calculated Fermi velocities as a function of twist angle ($\theta_M$) between top and bottom layers in FNTs (structure model shown in upper left inset). The velocity of collapsed (176,2) SWNT reduces to zero at which $\theta_M \approx 1.12°$. **d)** The inset on the bottom right presents the enlarged view of band structure from collapsed (176,2) SWNT. It shows moiré bands are very flat with bandwidth smaller than ~ 1 meV. **e)** Circumferential faceting of achiral double-walled nanotubes (carbon and boron nitride nanotubes). It tends to happen when the chiral angles between inner and outer nanotubes match, i.e., both inner and outer nanotubes are armchair (AC) or zigzag (ZZ). This kind of lattice reconstruction can impact the electronic properties of 1D moiré superlattices. a) Adapted with permission.[140] Copyright 2020, American Physical Society. b) Adapted with permission.[143] Copyright 2020, American Chemical Society. c) Adapted with permission.[134] Copyright 2015, American Physical Society. d) Adapted and reproduced with permission.[144] Copyright 2020, American Chemical Society. e) Adapted with permission.[162] Copyright 2016, Springer Nature Limited.

Type II band alignment is recently predicted to exist for large diameter DWNTs above a threshold diameter of 2.4 nm owing to the flexoelectricity effect between inner and outer nanotubes (Figure 6b).[143] This can potentially facilitate the formation of interlayer excitons and collection of photo-excited carriers in DWNTs and optoelectrical devices made from them. Heteronanotubes with different chemical compositions such as $MoS_2$/$WS_2$ can be intriguing systems to investigate moiré excitons in 1D as the study of 2D TMDC moiré superlattices have reveal rich and new physics.

### 4.2. Moiré flat bands

Can we realize 1D moiré flat bands with new correlated quantum phases in 1D moiré superlattices? Flat-band formation in DWNT moiré superlattices has been theoretically predicted as shown in Figure 6c due to the periodical moiré potential with long period.[134] In this example, the chiral vectors for inner (27,3) and outer (35,3) nanotubes are nearly parallel to each other with their difference orientating along the zigzag direction, which is a general criterion to generate the flat-band case in DWNT moiré superlattices. In such flat-band case, the on-site and long-range electron-electron Coulomb interaction can dominate over the kinetic energy such that insulating behavior at half- and fractional filling can occur, forming the Mott and generalized Wigner crystal states in 1D, like the case in 2D MATBG. Very interestingly, similar magic-angles to have flat moiré bands and vanishing Fermi velocities are recently predicted in another 1D moiré system, flattened carbon nanotubes (FNTs) as in 2D TBG (Figure 6d).[144] This hypothesized structure is transformed from a collapsed large-diameter nearly-zigzag SWNT (structure model shown in the upper left inset of Figure 6d). Notably, FNTs that are collapsed from nearly-zigzag nanotubes can even have narrower bandwidths (~ 1 meV) than MATBG does (inset on the bottom right of Figure 6d), which potentially makes it a stronger correlated electronic system. Experimentally, FNT samples can be produced by either unzipping of DWNTs[149,150] or extraction of inner shells of MWNTs[151], being the former with disconnected edge structures. 1D superconductivity in van der Waals materials may be possibly discovered in 1D moiré superlattices when changing the doping within the moiré flat bands. Moreover, electrons confined in 1D generally exhibits the well-known Luttinger liquid behavior where the motion of electrons are strongly correlated and only collective excitations are in principle allowed.[152,153] 1D moiré superlattices may provide a unique system to investigate the interplay of the two problems of electron correlation from both experiments and theoretical perspectives.

Electrical transport will be the most suitable technique to experimentally probe the correlated phenomena in 1D moiré superlattices near the Fermi energy. A critical component in the measurement is the ability to fabricate ultraclean samples and electrical devices. However, the hBN capping technique that is widely used in fabricating high-quality 2D heterostructures cannot be easily transferred to 1D. People have successfully worked out a device configuration in which nanotube samples were directly grown on pre-patterned electrodes.[154,155] Combing the direct growth and vdW assembly technique is an alternative and practically useful route to fabricate clean devices made of 1D structures.[156,157] An even bigger challenge is to find the target sample showing flat-band physics which is caused by the inability to controlling 1D vdW superlattices in the sample growth.

### 4.3. Strain and structural reconstruction

What is influence of strain and lattice reconstruction on electronic properties of 1D moiré superlattices? The present research of 2D moiré superlattices suffers from the sample inhomogeneity across different domains.[158,159] The variation of twisting angles over micrometer scale is believed to come from the non-uniform stain induced in the vdW stacking process,[160,161] which is very difficult to be satisfactorily solved. One intriguing possibility in 1D moiré is that it may allow higher homogeneity of moiré period over micrometer length scale since they are prepared by a direct growth technique. This has been partially supported by electron beam diffraction data of long suspended DWNTs.[89,119] On the other hand, lattice relaxation driven by the competing interlayer van der Waals coupling gain and intralayer strain cost may also happen in 1D moiré superlattices. People have found effect of circumferential faceting[162-164] in carbon (middle row) and boron nitride (bottom row) nanotubes when chiral angles match in different nanotube shells (Figure 6e).[162] Lattice reconstruction and its effect on the electronic properties have been recently noticed in 2D moiré systems.[165-169] However, so far, little is known for lattice reconstruction in 1D moiré superlattices.

### 4.4. Characterization tools

Developing convenient tools to directly characterize 1D moiré superlattices is the key to rapidly expand this research field and to correlate the measured properties with the moiré structure. Scanning probe microscopy with high throughput such as scanning electron microscopy (SEM),[170] piezoelectric force microscopy (PFM),[171] conductive atomic force microscopy (c-AFM)[172] and scanning near-field optical microscopy (SNOM)[173] are available choices which have been exploited in characterizing the 2D counterparts. Direct virtualization of the moiré flat bands in k space was recently achieved in 2D moiré superlattices with the advent of nano-ARPES.[58,59] Their applications to 1D moiré superlattices can be very important.

### 4.5. Expansion of 1D moiré systems

Besides DWNTs that have been extensively discussed, boron nitride nanotubes,[73,74] FNTs,[149-151] synthetic 1D heterostructures[77] and 2D heterostructures exhibiting anisotropy[156,174,175] are on the accessible material list of 1D moiré superlattices. The synthetic approach of 1D heterostructures are displayed in Figure 7a[176]. So far, synthesis of carbon/hBN, carbon/$MoS_2$ (Figures 7b-7d) and carbon/hBN/$MoS_2$ heteronanotubes have been successfully demonstrated,[77] greatly expanding the members of 1D vdW moiré superlattices. It promises the experimental realization of multi-walled nanotubes composed of more than two tubes of any materials, just like the famous "lego" in 2D.[177] But control of atomic registry between adjacent nanotubes are not possible at this point. Also the lengths of heteronanotubes (except for the carbon/hBN) are limited within few hundreds of nm due to the very slow growth rate of the outer tube via the inner tube template.[176] Growing longer heteronanotubes and investigating the growth mechanism are important directions for future research and applications.

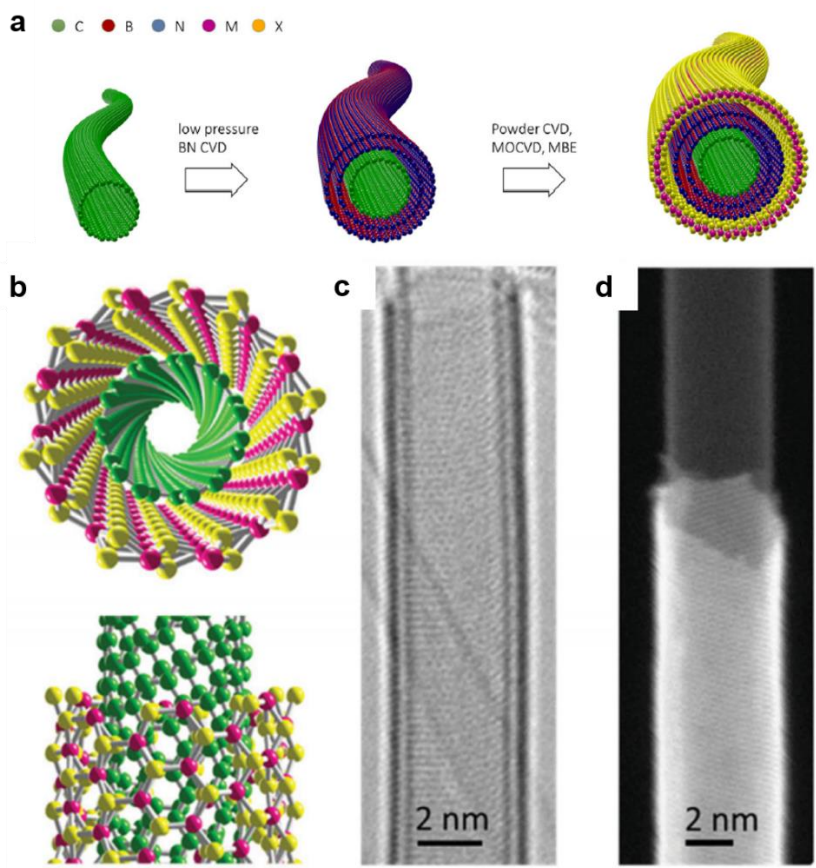

**Figure 7. Synthesis of 1D vdW heterostructures. a)** Scheme of the proposed growth approach of 1D vdW heterostructures. Nanotubes comprising of a large variety of chemical compositions may be realized via the layer-by-layer growth. **b, c, d)** Structure models, high resolution TEM (HRTEM image) and high-angle annular dark field (HAADF) scanning-TEM image of a single-walled $MoS_2$ nanotube grown on a SWNT. a) Reproduced with permission.[176] Copyright 2021, Wiley-VCH. b-d) Reproduced with permission.[77] Copyright 2020, American Association for the Advancement of Science.

The study of 1D van der Waals moiré superlattices is in its early stage. There are many exciting phenomena to be demonstrated, such as the evidence of moiré flat bands and correlated physics.

Meanwhile, experimental challenges on fabricating ultraclean devices and identifying the suitable type of 1D moiré superlattices showing strongly correlated behavior are waited to be overcome. We envision research attention on 1D van der Waals moiré superlattices will continue to increase, with substantial amount of efforts and works to be devoted to this emerging field.

## Acknowledgements

We thank Prof. Kazunari Matsuda and Prof. Yuhei Miyauchi in Kyoto University for their helpful discussion. S.Z. was supported by ZJU 100 Talents Program (107200*1942221R3) and (107200*194212101). R.K. was supported by JST CREST JPMJCR16F3 and JPMJCR19H4, and JST PRESTO JPMJPR20A2. P.M. acknowledges the support by National Science Foundation of China (Grant No. 12074260), Science and Technology Commission of Shanghai Municipality (Shanghai Natural Science Grants, Grant No. 19ZR1436400), and the NYU-ECNU Institute of Physics at NYU Shanghai. M.K. was supported by JSPS KAKENHI Grant Numbers JP20H01840 and JP20H00127, and by JST CREST Grant Number JPMJCR20T3, Japan.

## Competing interests

The authors declare no competing interests.

## Author contributions

All authors participated in researching the data, discussing the content and contributing to the writing of the manuscript.

**References:**

[1] A. D. Yoffe, *Annu. Rev. Mater. Sci.* **1993**, *3*, 147.
[2] F. Kelley, *Physics of Graphite,* Elsevier, New York **1981**.
[3] K. S. Novoselov, A. K. Geim, S. V. Morozov, D. Jiang, Y. Zhang, S. V. Dubonos, I. V. Grigorieva, A. A. Firsov, *Science* **2004**, *306*, 666.
[4] A. K. Geim, K. S. Novoselov, *Nat. Mater.* **2007**, *6*, 183.
[5] Q. H. Wang, K. Kalantar-Zadeh, A. Kis, J. N. Coleman, M. S. Strano, *Nat. Nanotechnol.* **2012**, *7*, 699.
[6] A. K. Geim, I. V. Grigorieva, *Nature* **2013**, *499*, 419.
[7] K. S. Novoselov, A. Mishchenko, A. Carvalho, A. H. Castro Neto, *Science* **2016**, *353*.
[8] E. McCann, M. Koshino, *Rep. Prog. Phys.* **2013**, *76*, 056503.
[9] Y. Zhang, T. Tang, C. Girit, Z. Hao, M. C. Martin, A. Zettl, M. F. Crommie, Y. R. Shen, F. Wang, *Nature* **2009**, *459*, 820.
[10] Jr. J. Velasco, L. Jing, W. Bao, Y. Lee, P. Kratz, V. Aji, M. Bockrath, C. N. Lau, C. Varma, R. Stillwell, D. Smirnov, F. Zhang, J. Jung, A. H. MacDonald, *Nat. Nanotechnol.* **2012**, *7*, 156.
[11] J. W. Mcclure, *Carbon* **1969**, *7*, 425.
[12] M. Koshino, E. McCann, *Phys. Rev. B* **2009**, *80*, 165409.
[13] W. Bao, L. Jing, Jr. J. Velasco, Y. Lee, G. Liu, D. Tran, B. Standley, M. Aykol, S. B. Cronin, D. Smirnov, M. Koshino, E. McCann, M. Bockrath, C. N. Lau, *Nat. Phys.* **2011**, *7*, 948.
[14] A. H. Castro Neto, F. Guinea, N. M. R. Peres, K. S. Novoselov, A. K. Geim, *Rev. Mod. Phys.* **2009**, *81*, 109.
[15] K. S. Novoselov, A. K. Geim, S. V. Morozov, D. Jiang, M. I. Katsnelson, I. V. Grigorieva, S. V. Dubonos, A. A. Firsov, *Nature* **2005**, *438*, 197.
[16] Y. B. Zhang, Y. W. Tan, H. L. Stormer, P. Kim, *Nature* **2005**, *438*, 201.
[17] K. F. Mak, C. Lee, J. Hone, J. Shan, T. F. Heinz, *Phys. Rev. Lett.* **2010**, *105*, 136805.
[18] A. Splendiani, L. Sun, Y. Zhang, T. Li, J. Kim, C. Chim, G. Galli, F. Wang, *Nano Lett.* **2010**, *10*, 1271.
[19] D. Xiao, G. Liu, W. Feng, X. Xu, W. Yao, *Phys. Rev. Lett.* **2012**, *108*, 196802.
[20] B. Huang, G. Clark, E. Navarro-Moratalla, D. R. Klein, R. Cheng, K. L. Seyler, D. Zhong, E. Schmidgall, M. A. McGuire, D. H. Cobden, W. Yao, D. Xiao, P. Jarillo-Herrero, X. Xu, *Nature* **2017**, *546*, 270.
[21] C. Gong, X. Zhang, *Science* **2019**, *363*, 706.
[22] M. Gibertini, M. Koperski, A. F. Morpurgo, K. S. Novoselov, *Nat. Nanotechnol.* **2019**, *14*, 408.
[23] F. Wang, T. A. Shifa, P. Yu, P. He, Y. Liu, F. Wang, Z. Wang, X. Zhan, X. Lou, F. Xia, J. He, *Adv. Funct. Mater.* **2018**, *28*, 1802151.
[24] C. R. Dean, A. F. Young, I. Meric, C. Lee, L. Wang, S. Sorgenfrei, K. Watanabe, T. Taniguchi, P. Kim, K. L. Shepard, J. Hone, *Nat. Nanotechnol.* **2010**, *5*, 722.
[25] L. Wang, I. Meric, P. Y. Huang, Q. Gao, Y. Gao, H. Tran, T. Taniguchi, K. Watanabe, L. M. Campos, D. A. Muller, J. Guo, P. Kim, J. Hone, K. L. Shepard, C. R. Dean, *Science* **2013**, *342*, 614.

[26] J. M. B. L. Dos Santos, N. M. R. Peres, A. H. Castro Neto, *Phys. Rev. Lett.* **2007**, *99*, 256802.
[27] E. Suarez Morell, J. D. Correa, P. Vargas, M. Pacheco, Z. Barticevic, *Phys. Rev. B* **2010**, *82*, 121407.
[28] G. Li, A. Luican, J. M. B. Lopes Dos Santos, A. H. Castro Neto, A. Reina, J. Kong, E. Y. Andrei, *Nat. Phys.* **2010**, *6*, 109.
[29] R. Bistritzer, A. H. MacDonald, *Proc. Natl. Acad. Sci. U.S.A* **2011**, *108*, 12233.
[30] S. Carr, D. Massatt, S. Fang, P. Cazeaux, M. Luskin, E. Kaxiras, *Phys. Rev. B* **2017**, *95*, 075420.
[31] R. Ribeiro-Palau, C. Zhang, K. Watanabe, T. Taniguchi, J. Hone, C. R. Dean, *Science* **2018**, *361*, 690.
[32] Y. Cao, V. Fatemi, A. Demir, S. Fang, S. L. Tomarken, J. Y. Luo, J. D. Sanchez-Yamagishi, K. Watanabe, T. Taniguchi, E. Kaxiras, R. C. Ashoori, P. Jarillo-Herrero, *Nature* **2018**, *556*, 80.
[33] Y. Cao, V. Fatemi, S. Fang, K. Watanabe, T. Taniguchi, E. Kaxiras, P. Jarillo-Herrero, *Nature* **2018**, *556*, 43.
[34] E. Y. Andrei, A. H. MacDonald, *Nat. Mater.* **2021**, *20*, 571.
[35] W. Yao, M. Aeschlimann, S. Zhou, *Chinese Phys. B* **2020**, *29*, 127304.
[36] L. Wang, Y. Gao, B. Wen, Z. Han, T. Taniguchi, K. Watanabe, M. Koshino, J. Hone, C. R. Dean, *Science* **2015**, *350*, 1231.
[37] O. A. Ajayi, J. V. Ardelean, G. D. Shepard, J. Wang, A. Antony, T. Taniguchi, K. Watanabe, T. F. Heinz, S. Strauf, X. Zhu, J. C. Hone, *2D Mater* **2017**, *4*, 031011.
[38] G. X. Ni, A. S. McLeod, Z. Sun, L. Wang, L. Xiong, K. W. Post, S. S. Sunku, B. Jiang, J. Hone, C. R. Dean, M. M. Fogler, D. N. Basov, *Nature* **2018**, *557*, 530.
[39] M. Yankowitz, Q. Ma, P. Jarillo-Herrero, B. J. LeRoy, *Nat. Rev. Phys.* **2019**, *1*, 112.
[40] M. Yankowitz, J. Jung, E. Laksono, N. Leconte, B. L. Chittari, K. Watanabe, T. Taniguchi, S. Adam, D. Graf, C. R. Dean, *Nature* **2018**, *557*, 404.
[41] X. Lu, P. Stepanov, W. Yang, M. Xie, M. A. Aamir, I. Das, C. Urgell, K. Watanabe, T. Taniguchi, G. Zhang, A. Bachtold, A. H. MacDonald, D. K. Efetov, *Nature* **2019**, *574*, 653.
[42] Y. Xie, B. Lian, B. Jack, X. Liu, C. Chiu, K. Watanabe, T. Taniguchi, B. A. Bernevig, A. Yazdani, *Nature* **2019**, *572*, 101.
[43] G. Chen, L. Jiang, S. Wu, B. Lyu, H. Li, B. L. Chittari, K. Watanabe, T. Taniguchi, Z. Shi, J. Jung, Y. Zhang, F. Wang, *Nat. Phys.* **2019**, *15*, 237.
[44] G. Chen, A. L. Sharpe, P. Gallagher, I. T. Rosen, E. J. Fox, L. Jiang, B. Lyu, H. Li, K. Watanabe, T. Taniguchi, J. Jung, Z. Shi, D. Goldhaber-Gordon, Y. Zhang, F. Wang, *Nature* **2019**, *572*, 215.
[45] K. L. Seyler, P. Rivera, H. Yu, N. P. Wilson, E. L. Ray, D. G. Mandrus, J. Yan, W. Yao, X. Xu, *Nature* **2019**, *567*, 66.
[46] K. Tran, G. Moody, F. Wu, X. Lu, J. Choi, K. Kim, A. Rai, D. A. Sanchez, J. Quan, A. Singh, J. Embley, A. Zepeda, M. Campbell, T. Autry, T. Taniguchi, K. Watanabe, N. Lu, S. K. Banerjee, K. L. Silverman, S. Kim, E. Tutuc, L. Yang, A. H. MacDonald, X. Li, *Nature* **2019**, *567*, 71.
[47] C. Jin, E. C. Regan, A. Yan, M. I. B. Utama, D. Wang, S. Zhao, Y. Qin, S. Yang, Z. Zheng, S. Shi, K. Watanabe, T. Taniguchi, S. Tongay, A. Zettl, F. Wang, *Nature* **2019**, *567*, 76.

[48]    E. M. Alexeev, D. A. Ruiz-Tijerina, M. Danovich, M. J. Hamer, D. J. Terry, P. K. Nayak, S. Ahn, S. Pak, J. Lee, J. I. Sohn, M. R. Molas, M. Koperski, K. Watanabe, T. Taniguchi, K. S. Novoselov, R. V. Gorbachev, H. S. Shin, V. I. Fal'Ko, A. I. Tartakovskii, *Nature* **2019**, *567*, 81.
[49]    E. C. Regan, D. Wang, C. Jin, M. I. Utama, B. Gao, X. Wei, S. Zhao, W. Zhao, Z. Zhang, K. Yumigeta, M. Blei, J. D. Carlstrom, K. Watanabe, T. Taniguchi, S. Tongay, M. Crommie, A. Zettl, F. Wang, *Nature* **2020**, *579*, 359.
[50]    Y. Tang, L. Li, T. Li, Y. Xu, S. Liu, K. Barmak, K. Watanabe, T. Taniguchi, A. H. MacDonald, J. Shan, K. F. Mak, *Nature* **2020**, *579*, 353.
[51]    Y. Xu, S. Liu, D. A. Rhodes, K. Watanabe, T. Taniguchi, J. Hone, V. Elser, K. F. Mak, J. Shan, *Nature* **2020**, *587*, 214.
[52]    Y. Shimazaki, I. Schwartz, K. Watanabe, T. Taniguchi, M. Kroner, A. Imamoglu, *Nature* **2020**, *580*, 472.
[53]    Y. Zhang, D. Mao, Y. Cao, P. Jarillo-Herrero, T. Senthil, *Phys. Rev. B* **2019**, *99*, 075127.
[54]    N. Bultinck, S. Chatterjee, M. P. Zaletel, *Phys. Rev. Lett.* **2020**, *124*, 166601.
[55]    G. Tarnopolsky, A. J. Kruchkov, A. Vishwanath, *Phys. Rev. Lett.* **2019**, *122*, 106405.
[56]    F. Wu, T. Lovorn, E. Tutuc, A. H. MacDonald, *Phys. Rev. Lett.* **2018**, *121*, 026402.
[57]    L. Balents, C. R. Dean, D. K. Efetov, A. F. Young, *Nat. Phys.* **2020**, *16*, 725.
[58]    M. I. B. Utama, R. J. Koch, K. Lee, N. Leconte, H. Li, S. Zhao, L. Jiang, J. Zhu, K. Watanabe, T. Taniguchi, P. D. Ashby, A. Weber-Bargioni, A. Zettl, C. Jozwiak, J. Jung, E. Rotenberg, A. Bostwick, F. Wang, *Nat. Phys.* **2021**, *17*, 184.
[59]    S. Lisi, X. Lu, T. Benschop, T. A. de Jong, P. Stepanov, J. R. Duran, F. Margot, I. Cucchi, E. Cappelli, A. Hunter, A. Tamai, V. Kandyba, A. Giampietri, A. Barinov, J. Jobst, V. Stalman, M. Leeuwenhoek, K. Watanabe, T. Taniguchi, L. Rademaker, S. J. van der Molen, M. P. Allan, D. K. Efetov, F. Baumberger, *Nat. Phys.* **2021**, *17*, 189.
[60]    A. L. Sharpe, E. J. Fox, A. W. Barnard, J. Finney, K. Watanabe, T. Taniguchi, M. A. Kastner, D. Goldhaber-Gordon, *Science* **2019**, *365*, 605.
[61]    M. Serlin, C. L. Tschirhart, H. Polshyn, Y. Zhang, J. Zhu, K. Watanabe, T. Taniguchi, L. Balents, A. F. Young, *Science* **2020**, *367*, 900.
[62]    H. Polshyn, J. Zhu, M. A. Kumar, Y. Zhang, F. Yang, C. L. Tschirhart, M. Serlin, K. Watanabe, T. Taniguchi, A. H. MacDonald, A. F. Young, *Nature* **2020**, *588*, 66.
[63]    G. Chen, A. L. Sharpe, E. J. Fox, Y. Zhang, S. Wang, L. Jiang, B. Lyu, H. Li, K. Watanabe, T. Taniguchi, Z. Shi, T. Senthil, D. Goldhaber-Gordon, Y. Zhang, F. Wang, *Nature* **2020**, *579*, 56.
[64]    S. Wu, Z. Zhang, K. Watanabe, T. Taniguchi, E. Y. Andrei, *Nat. Mater.* **2021**, *20*, 488.
[65]    D. Rodan-Legrain, Y. Cao, J. M. Park, S. C. de la Barrera, M. T. Randeria, K. Watanabe, T. Taniguchi, P. Jarillo-Herrero, *Nat. Nanotechnol.* **2021**.
[66]    F. K. de Vries, E. Portoles, G. Zheng, T. Taniguchi, K. Watanabe, T. Ihn, K. Ensslin, P. Rickhaus, *Nat. Nanotechnol.* **2021**.
[67]    C. Wood, *Science* **2020**, *370*, 897.
[68]    C. Shen, A. H. Brozena, Y. Wang, *Nanoscale* **2011**, *3*, 503.
[69]    Y. A. Kim, K. Yang, H. Muramatsu, T. Hayashi, M. Endo, M. Terrones, M. S. Dresselhaus, *Carbon Lett.* **2014**, *15*, 77.
[70]    K. E. Moore, D. D. Tune, B. S. Flavel, *Adv. Mater.* **2015**, *27*, 3105.

[71]   K. Fujisawa, H. J. Kim, S. H. Go, H. Muramatsu, T. Hayashi, M. Endo, T. C. Hirschmann, M. S. Dresselhaus, Y. A. Kim, P. T. Araujo, *Appl. Sci.* **2016**, *6*.
[72]   R. Andrews, D. Jacques, D. L. Qian, T. Rantell, *Acc. Chem. Res.* **2002**, *35*, 1008.
[73]   M. L. Cohen, A. Zettl, *Phys. Today* **2010**, *63*, 34.
[74]   D. Golberg, Y. Bando, Y. Huang, T. Terao, M. Mitome, C. Tang, C. Zhi, *ACS Nano* **2010**, *4*, 2979.
[75]   M. Nath, A. Govindaraj, C. Rao, *Adv. Mater.* **2001**, *13*, 283.
[76]   R. Nakanishi, R. Kitaura, J. H. Warner, Y. Yamamoto, S. Arai, Y. Miyata, H. Shinohara, *Sci. Rep.* **2013**, *3*, 1385.
[77]   R. Xiang, T. Inoue, Y. Zheng, A. Kumamoto, Y. Qian, Y. Sato, M. Liu, D. Tang, D. Gokhale, J. Guo, K. Hisama, S. Yotsumoto, T. Ogamoto, H. Arai, Y. Kobayashi, H. Zhang, B. Hou, A. Anisimov, M. Maruyama, Y. Miyata, S. Okada, S. Chiashi, Y. Li, J. Kong, E. I. Kauppinen, Y. Ikuhara, K. Suenaga, S. Maruyama, *Science* **2020**, *367*, 537.
[78]   R. Saito, G. Dresselhaus, & M. S. Dresselhaus, *Physical Properties of Carbon Nanotubes*, Imperial College Press **1998**.
[79]   R. B. Weisman, S. M. Bachilo, *Nano Lett.* **2003**, *3*, 1235.
[80]   M. S. Dresselhaus, G. Dresselhaus, R. Saito, A. Jorio, *Phys. Rep.* **2005**, *409*, 47.
[81]   K. Liu, J. Deslippe, F. Xiao, R. B. Capaz, X. Hong, S. Aloni, A. Zettl, W. Wang, X. Bai, S. G. Louie, E. Wang, F. Wang, *Nat. Nanotechnol.* **2012**, *7*, 325.
[82]   K. Hirahara, M. Kociak, S. Bandow, T. Nakahira, K. Itoh, Y. Saito, S. Iijima, *Phys. Rev. B* **2006**, *73*, 195420.
[83]   A. Ghedjatti, Y. Magnin, F. Fossard, G. Wang, H. Amara, E. Flahaut, J. Lauret, A. Loiseau, *ACS Nano* **2017**, *11*, 4840.
[84]   M. He, Z. Xu, D. Shang, X. Zhang, H. Zhang, D. Li, H. Jiang, E. Kauppinen, F. Ding, *Carbon* **2019**, *144*, 147.
[85]   K. Schouteden, A. Volodin, Z. Li, C. Van Haesendonck, *Carbon* **2013**, *61*, 379.
[86]   M. Kociak, K. Hirahara, K. Suenaga, S. Iijima, *Eur. Phys. J. B* **2003**, *32*, 457.
[87]   H. Deniz, A. Derbakova, L. Qin, *Ultramicroscopy* **2010**, *111*, 66.
[88]   K. Liu, Z. Xu, W. Wang, P. Gao, W. Fu, X. Bai, E. Wang, *J. Phys. D Appl. Phys.* **2009**, *42*, 125412.
[89]   S. Zhao, T. Kitagawa, Y. Miyauchi, K. Matsuda, H. Shinohara, R. Kitaura, *Nano Res* **2014**, *7*, 1548.
[90]   Z. J. Liu, L. C. Qin, *Chem. Phys. Lett.* **2005**, *408*, 75.
[91]   L. Qin, *Rep. Prog. Phys.* **2006**, *69*, 2761.
[92]   H. Jiang, A. G. Nasibulin, D. P. Brown, E. I. Kauppinen, *Carbon* **2007**, *45*, 662.
[93]   Y. K. Kwon, S. Saito, D. Tomanek, *Phys. Rev. B* **1998**, *58*, 13314.
[94]   S. Sanvito, Y. K. Kwon, D. Tomanek, C. J. Lambert, *Phys. Rev. Lett.* **2000**, *84*, 1974.
[95]   R. Moradian, S. Azadi, H. Refii-Tabar, *J. Phys. Condens. Mat.* **2007**, *19*, 176209.
[96]   V. Zolyomi, J. Koltai, A. Rusznyak, J. Kuerti, A. Gali, F. Simon, H. Kuzmany, A. Szabados, P. R. Surjan, *Phys. Rev. B* **2008**, *77*, 245403.
[97]   S. Uryu, T. Ando, *Phys. Rev. B* **2005**, *72*, 245403.
[98]   K. Liu, C. Jin, X. Hong, J. Kim, A. Zettl, E. Wang, F. Wang, *Nat. Phys.* **2014**, *10*, 737.
[99]   T. Hertel, A. Hagen, V. Talalaev, K. Arnold, F. Hennrich, M. Kappes, S. Rosenthal, J. McBride, H. Ulbricht, E. Flahaut, *Nano Lett.* **2005**, *5*, 511.

[100] H. Hirori, K. Matsuda, Y. Kanemitsu, *Phys. Rev. B* **2008**, *78*, 113409.
[101] D. Shimamoto, H. Muramatsu, T. Hayashi, Y. A. Kim, M. Endo, J. S. Park, R. Saito, M. Terrones, M. S. Dresselhaus, *Appl. Phys. Lett.* **2009**, *94*, 083106.
[102] D. I. Levshov, R. Parret, T. Huy-Nam, T. Michel, T. C. Thi, N. Van Chuc, R. Arenal, V. N. Popov, S. B. Rochal, J. Sauvajol, A. Zahab, M. Paillet, *Phys. Rev. B* **2017**, *96*, 195410.
[103] Y. Ohno, S. Iwasaki, Y. Murakami, S. Kishimoto, S. Maruyama, T. Mizutani, *Phys. Rev. B* **2006**, *73*, 235427.
[104] Y. Miyauchi, R. Saito, K. Sato, Y. Ohno, S. Iwasaki, T. Mizutani, J. Jiang, S. Maruyama, *Chem. Phys. Lett.* **2007**, *442*, 394.
[105] K. Sato, R. Saito, J. Jiang, G. Dresselhaus, M. S. Dresselhaus, *Phys. Rev. B* **2007**, *76*, 195446.
[106] T. Ando, *J. Phys. Soc. Jpn* **2010**, *79*, 024706.
[107] Y. Tomio, H. Suzuura, T. Ando, *Phys. Rev. B* **2012**, *85*, 085411.
[108] M. S. Dresselhaus, A. Jorio, M. Hofmann, G. Dresselhaus, R. Saito, *Nano Lett.* **2010**, *10*, 751.
[109] A. C. Ferrari, J. C. Meyer, V. Scardaci, C. Casiraghi, M. Lazzeri, F. Mauri, S. Piscanec, D. Jiang, K. S. Novoselov, S. Roth, A. K. Geim, *Phys. Rev. Lett.* **2006**, *97*, 187401.
[110] F. Villalpando-Paez, H. Son, D. Nezich, Y. P. Hsieh, J. Kong, Y. A. Kim, D. Shimamoto, H. Muramatsu, T. Hayashi, M. Endo, M. Terrones, M. S. Dresselhaus, *Nano Lett.* **2008**, *8*, 3879.
[111] T. C. Hirschmann, P. T. Araujo, H. Muramatsu, X. Zhang, K. Nielsch, Y. A. Kim, M. S. Dresselhaus, *ACS Nano* **2013**, *7*, 2381.
[112] D. Levshov, T. X. Than, R. Arenal, V. N. Popov, R. Parret, M. Paillet, V. Jourdain, A. A. Zahab, T. Michel, Y. I. Yuzyuk, J. Sauvajol, *Nano Lett.* **2011**, *11*, 4800.
[113] K. Liu, X. Hong, M. Wu, F. Xiao, W. Wang, X. Bai, J. W. Ager, S. Aloni, A. Zettl, E. Wang, F. Wang, *Nat. Commun.* **2013**, *4*, 1375.
[114] V. N. Popov, L. Henrard, *Phys. Rev. B* **2002**, *65*, 235415.
[115] H. N. Tran, J. C. Blancon, R. Arenal, R. Parret, A. A. Zahab, A. Ayari, F. Vallee, N. Del Fatti, J. L. Sauvajol, M. Paillet, *Phys. Rev. B* **2017**, *95*, 205411.
[116] G. Gordeev, S. Waßerroth, H. Li, B. Flavel & S. Reich, (Preprint) arXiv:2103.07337, v1; submitted: Mar 2021.
[117] M. Kociak, K. Suenaga, K. Hirahara, Y. Saito, T. Nakahira, S. Iijima, *Phys. Rev. Lett.* **2002**, *89*, 155501.
[118] R. Bonnet, A. Lherbier, C. Barraud, M. L. Della Rocca, P. Lafarge, J. Charlier, *Sci. Rep.* **2016**, *6*, 19701.
[119] K. Liu, W. Wang, Z. Xu, X. Bai, E. Wang, Y. Yao, J. Zhang, Z. Liu, *J. Am. Chem. Soc.* **2009**, *131*, 62.
[120] K. E. Moore, M. Pfohl, D. D. Tune, F. Hennrich, S. Dehm, V. S. K. Chakradhanula, C. Kuebel, R. Krupke, B. S. Flavel, *ACS Nano* **2015**, *9*, 3849.
[121] D. Bouilly, J. Cabana, F. Meunier, M. Desjardins-Carriere, F. Lapointe, P. Gagnon, F. L. Larouche, E. Adam, M. Paillet, R. Martel, *ACS Nano* **2011**, *5*, 4927.
[122] K. Fujisawa, K. Komiyama, H. Muramatsu, D. Shimamoto, T. Tojo, Y. A. Kim, T. Hayashi, M. Endo, K. Oshida, M. Terrones, M. S. Dresselhaus, *ACS Nano* **2011**, *5*, 7547.

[123] M. Y. Sfeir, F. Wang, L. M. Huang, C. C. Chuang, J. Hone, S. P. O'Brien, T. F. Heinz, L. E. Brus, *Science* **2004**, *306*, 1540.
[124] M. Y. Sfeir, T. Beetz, F. Wang, L. M. Huang, X. Huang, M. Y. Huang, J. Hone, S. O'Brien, J. A. Misewich, T. F. Heinz, L. J. Wu, Y. M. Zhu, L. E. Brus, *Science* **2006**, *312*, 554.
[125] M. F. Islam, D. E. Milkie, C. L. Kane, A. G. Yodh, J. M. Kikkawa, *Phys. Rev. Lett.* **2004**, *93*, 037404.
[126] Y. Murakami, E. Einarsson, T. Edamura, S. Maruyama, *Phys. Rev. Lett.* **2005**, *94*, 087402.
[127] F. Vialla, C. Roquelet, B. Langlois, G. Delport, S. M. Santos, E. Deleporte, P. Roussignol, C. Delalande, C. Voisin, J. Lauret, *Phys. Rev. Lett.* **2013**, *111*, 137402.
[128] S. Berciaud, L. Cognet, P. Poulin, R. B. Weisman, B. Lounis, *Nano Lett.* **2007**, *7*, 1203.
[129] Blancon, M. Paillet, H. N. Tran, T. T. Xuan, S. A. Guebrou, A. Ayari, A. San Miguel, P. Ngoc-Minh, A. Zahab, J. Sauvajol, N. Del Fatti, F. Vallee, *Nat Commun* **2013**, *4*, 2542.
[130] K. Liu, X. Hong, S. Choi, C. Jin, R. B. Capaz, J. Kim, W. Wang, X. Bai, S. G. Louie, E. Wang, F. Wang, *P. Proc. Natl. Acad. Sci. U.S.A.* **2014**, *111*, 7564.
[131] Maruyama, R. Kojima, Y. Miyauchi, S. Chiashi, M. Kohno, *Chem. Phys. Lett.* **2002**, *360*, 229.
[132] A. Ishii, M. Yoshida, Y. K. Kato, *Phys. Rev. B* **2015**, *91*, 125427.
[133] S. D. Wang, M. Grifoni, *Phys. Rev. Lett.* **2005**, *95*, 266802.
[134] M. Koshino, P. Moon, Y. Son, *Phys. Rev. B* **2015**, *91*, 035405.
[135] S. Shallcross, S. Sharma, O. A. Pankratov, *Phys. Rev. Lett.* **2008**, *101*, 056803.
[136] J. R. Wallbank, A. A. Patel, M. Mucha-Kruczynski, A. K. Geim, V. I. Falko, *Phys. Rev. B* **2013**, *87*, 245408.
[137] P. Moon, M. Koshino, *Phys. Rev. B* **2013**, *87*, 205404.
[138] F. Gargiulo, O. V. Yazyev, *2D Mater.* **2018**, *5*, 015019.
[139] S. Zhao, P. Moon, Y. Miyauchi, T. Nishihara, K. Matsuda, M. Koshino, R. Kitaura, *Phys. Rev. Lett.* **2020**, *124*, 106101.
[140] D. V. Chalin, S. B. Rochal, *Phys. Rev. B* **2020**, *102*, 115426.
[141] V. N. Popov, *Carbon* **2020**, *170*, 30.
[142] Y. Su, S. Lin, *Phys. Rev. B* **2020**, *101*, 041113(R).
[143] V. I. Artyukhov, S. Gupta, A. Kutana, B. I. Yakobson, *Nano Lett.* **2020**, *20*, 3240.
[144] O. Arroyo-Gascon, R. Fernandez-Perea, E. Suarez Morell, C. Cabrillo, L. Chico, *Nano Lett.* **2020**, *20*, 7588.
[145] H. Yu, G. Liu, J. Tang, X. Xu, W. Yao, *Sci. Adv.* **2017**, *3*, e1701696.
[146] C. Zhang, C. Chuu, X. Ren, M. Li, L. Li, C. Jin, M. Chou, C. Shih, *Sci. Adv.* **2017**, *3*, e1601459.
[147] C. Jin, E. C. Regan, D. Wang, M. I. B. Utama, C. Yang, J. Cain, Y. Qin, Y. Shen, Z. Zheng, K. Watanabe, T. Taniguchi, S. Tongay, A. Zettl, F. Wang, *Nat. Phys.* **2019**, *15*, 1140.
[148] H. Baek, M. Brotons-Gisbert, Z. X. Koong, A. Campbell, M. Rambach, K. Watanabe, T. Taniguchi, B. D. Gerardot, *Sci. Adv.* **2020**, *6*, eaba8526.
[149] D. V. Kosynkin, A. L. Higginbotham, A. Sinitskii, J. R. Lomeda, A. Dimiev, B. K. Price, J. M. Tour, *Nature* **2009**, *458*, 872.
[150] L. Jiao, L. Zhang, X. Wang, G. Diankov, H. Dai, *Nature* **2009**, *458*, 877.
[151] D. H. Choi, Q. Wang, Y. Azuma, Y. Majima, J. H. Warner, Y. Miyata, H. Shinohara, R. Kitaura, *Sci. Rep.* **2013**, *3*, 1617.

[152] T. Giamarchi, *Quantum Physics in One Dimension,* Oxford University Press **2004**.
[153] S. Zhao, S. Wang, F. Wu, W. Shi, I. B. Utama, T. Lyu, L. Jiang, Y. Su, S. Wang, K. Watanabe, T. Taniguchi, A. Zettl, X. Zhang, C. Zhou, F. Wang, *Phys. Rev. Lett.* **2018**, *121*, 047702.
[154] V. V. Deshpande, B. Chandra, R. Caldwell, D. S. Novikov, J. Hone, M. Bockrath, *Science* **2009**, *323*, 106.
[155] I. Shapir, A. Hamo, S. Pecker, C. P. Moca, O. Legeza, G. Zarand, S. Ilani, *Science* **2019**, *364*, 870.
[156] S. Zhao, S. Yoo, S. Wang, B. Lyu, S. Kahn, F. Wu, Z. Zhao, D. Cui, W. Zhao, Y. Yoon, M. I. B. Utama, W. Shi, K. Watanabe, T. Taniguchi, M. F. Crommie, Z. Shi, C. Zhou, F. Wang, *Nano Lett.* **2020**, *20*, 6712.
[157] K. Otsuka, N. Fang, D. Yamashita, T. Taniguchi, K. Watanabe, Y. K. Kato, *Nat. Commun.* **2021**, *12*, 3138.
[158] Y. Bai, L. Zhou, J. Wang, W. Wu, L. J. McGilly, D. Halbertal, C. F. B. Lo, F. Liu, J. Ardelean, P. Rivera, N. R. Finney, X. Yang, D. N. Basov, W. Yao, X. Xu, J. Hone, A. N. Pasupathy, X. Zhu, *Nat. Mater.* **2020**, *19*, 1068.
[159] A. Uri, S. Grover, Y. Cao, J. A. Crosse, K. Bagani, D. Rodan-Legrain, Y. Myasoedov, K. Watanabe, T. Taniguchi, P. Moon, M. Koshino, P. Jarillo-Herrero, E. Zeldov, *Nature* **2020**, *581*, 47.
[160] K. Kim, M. Yankowitz, B. Fallahazad, S. Kang, H. C. P. Movva, S. Huang, S. Larentis, C. M. Corbet, T. Taniguchi, K. Watanabe, S. K. Banerjee, B. J. LeRoy, E. Tutuc, *Nano Lett.* **2016**, *16*, 1989.
[161] D. G. Purdie, N. M. Pugno, T. Taniguchi, K. Watanabe, A. C. Ferrari, A. Lombardo, *Nat. Commun.* **2018**, *9*.
[162] I. Leven, R. Guerra, A. Vanossi, E. Tosatti, O. Hod, *Nat. Nanotechnol.* **2016**, *11*, 1082.
[163] Y. Gogotsi, J. A. Libera, N. Kalashnikov, M. Yoshimura, *Science* **2000**, *290*, 317.
[164] D. Golberg, M. Mitome, Y. Bando, C. C. Tang, C. Y. Zhi, *Appl. Phys. A-Mater.* **2007**, *88*, 347.
[165] C. R. Woods, L. Britnell, A. Eckmann, R. S. Ma, J. C. Lu, H. M. Guo, X. Lin, G. L. Yu, Y. Cao, R. V. Gorbachev, A. V. Kretinin, J. Park, L. A. Ponomarenko, M. I. Katsnelson, Y. N. Gornostyrev, K. Watanabe, T. Taniguchi, C. Casiraghi, H. Gao, A. K. Geim, K. S. Novoselov, *Nat. Phys.* **2014**, *10*, 451.
[166] H. Yoo, R. Engelke, S. Carr, S. Fang, K. Zhang, P. Cazeaux, S. H. Sung, R. Hoyden, A. W. Tsen, T. Taniguchi, K. Watanabe, G. Yi, M. Kim, M. Luskin, E. B. Tadmor, E. Kaxiras, P. Kim, *Nat. Mater.* **2019**, *18*, 448.
[167] A. Weston, Y. Zou, V. Enaldiev, A. Summerfield, N. Clark, V. Zolyomi, A. Graham, C. Yelgel, S. Magorrian, M. Zhou, J. Zultak, D. Hopkinson, A. Barinov, T. H. Bointon, A. Kretinin, N. R. Wilsons, P. H. Beton, V. I. Fal'Ko, S. J. Haigh, R. Gorbachev, *Nat. Nanotechnol.* **2020**, *15*, 592.
[168] H. Li, S. Li, M. H. Naik, J. Xie, X. Li, J. Wang, E. Regan, D. Wang, W. Zhao, S. Zhao, S. Kahn, K. Yumigeta, M. Blei, T. Taniguchi, K. Watanabe, S. Tongay, A. Zettl, S. G. Louie, F. Wang, M. F. Crommie, *Nat. Mater.* **2021**, *20*, 945.
[169] N. N. T. Nam, M. Koshino, *Phys. Rev. B* **2020**, *101*, 099901.

[170] T. I. Andersen, G. Scuri, A. Sushko, K. De Greve, J. Sung, Y. Zhou, D. S. Wild, R. J. Gelly, H. Heo, D. Berube, A. Y. I. Joe, L. A. Jauregui, K. Watanabe, T. Taniguchi, P. Kim, H. Park, M. D. Lukin, *Nat. Mater.* **2021**, *20*, 480.

[171] L. J. McGilly, A. Kerelsky, N. R. Finney, K. Shapovalov, E. Shih, A. Ghiotto, Y. Zeng, S. L. Moore, W. Wu, Y. Bai, K. Watanabe, T. Taniguchi, M. Stengel, L. Zhou, J. Hone, X. Zhu, D. N. Basov, C. Dean, C. E. Dreyer, A. N. Pasupathy, *Nat. Nanotechnol.* **2020**, *15*, 580.

[172] S. Zhang, A. Song, L. Chen, C. Jiang, C. Chen, L. Gao, Y. Hou, L. Liu, T. Ma, H. Wang, X. Feng, Q. Li, *Sci. Adv.* **2020**, *6*, eabc5555.

[173] H. Li, M. I. B. Utama, S. Wang, W. Zhao, S. Zhao, X. Xiao, Y. Jiang, L. Jiang, T. Taniguchi, K. Watanabe, A. Weber-Bargioni, A. Zettl, F. Wang, *Nano Lett.* **2020**, *20*, 3106.

[174] D. M. Kennes, L. Xian, M. Claassen, A. Rubio, *Nat. Commun.* **2020**, *11*, 1124.

[175] J. Mao, S. P. Milovanovic, M. Andelkovic, X. Lai, Y. Cao, K. Watanabe, T. Taniguchi, L. Covaci, F. M. Peeters, A. K. Geim, Y. Jiang, E. Y. Andrei, *Nature* **2020**, *584*, 215.

[176] R. Xiang & S. Maruyama, *Small Sci.* **2021**, 2000039.

[177] P. Ajayan, P. Kim, K. Banerjee, *Phys. Today* **2016**, *69*, 39.